\documentclass[10pt]{article}
\begin{document}
{\begin{center}
{\LARGE \bf Dynamical symmetries and well-localized  \\

\vspace{0.2cm}

hydrogenic wave packets \footnote{Submitted to Proceedings of Institute of 
Mathematics of NAS of Ukraine.}}

\vspace{0.7cm}
Vladimir ZVEREV~$^\dag$ and Boris RUBINSTEIN~$^\ddag$

\vspace{0.2cm}
\textit{~$^\dag$~Ural State Technical University, Ekaterinburg, 620002, Russia}

\textit{zverev@dpt.ustu.ru}

\vspace{0.2cm}
\textit{~$^\ddag$~Department of Mathematics, University of California, 
Davis, CA 95616, USA}

\textit{boris@math.ucdavis.edu}

\end{center}}
\vspace{0.2cm}

\begin{abstract}
A new method for constructing of composite
coherent states of the hydrogen atom, based on the dynamical group
approach and various schemes of reduction to subgroups, 
is presented. A wide class of well-localized (Gaussian) hydrogenic wave
packets for
circular and elliptic orbits is found using the saddle-point
method.
\end{abstract}

\section{Introduction}

In recent years, new experimental techniques opened way to creation and
study of high energy (Rydberg) states in atoms. These states are
described by approximate hydrogenic wave functions with very large principal
quantum numbers. Some new effects, as the dynamical localization and the
dynamical chaos, have attracted considerable interest. Explanation of these
phenomena uses classical equations of motion \cite{Zverev:bellomo&stroud&%
farelly}. It is reasonable to look for an alternative quantum description on 
the basis of semi-classical approximations, which is naturally provided by a 
coherent states (CS) formalism.

In Section 2, starting from the $\mbox{O}(4,2)$ dynamical group approach 
\cite{Zverev:barut&raczka} and using three schemes of reduction to
subgroups \cite{Zverev:zverev&rubinstein}: $\mbox{O}(4,2) \supset \mbox{O}(4) 
\sim \mbox{O}(3) \otimes \mbox{O}(3), \mbox{O}(4,2)\supset \mbox{O}(2,2) 
\sim \mbox{O}(2,1) \otimes \mbox{O}(2,1), \mbox{O}(4,2) \supset \mbox{O}(3) 
\otimes \mbox{O}(2,1)$, we construct composite CS in physical and auxiliary 
("tilted") representations \cite{Zverev:bechler}. We use two types of 
generating operators of CS with different procedures of transition to a 
classical limit. In particular, the generating operators for Perelomov 
$\mbox{SO}(3)$ and $\mbox{SO}(2,1)$ CS \cite{Zverev:perelomov}, 
Barut-Girardello $\mbox{SO}(2,1)$ CS \cite{Zverev:barut&raczka,Zverev:barut%
&girardello}, generalized hypergeometric CS \cite{Zverev:zverev}, Brif 
$\mbox{SO}(3)$ and $\mbox{SO}(2,1)$ algebra eigenstates \cite{Zverev:brif}, 
may be used for this purpose. The CS are separated into \textit{two
classes} with different semi-classical behavior.

The hydrogenic CS wave functions have complicated form, so it is reasonable
to use simplified asymptotic expressions. In Section 3 we describe a method
for asymptotic estimate and obtain well-localized hydrogenic wave packets for 
circular and elliptic orbits. A similar asymptotic estimate method is used 
in the theory of CS path integrals. We believe that the approach discussed 
in this paper can be applied to computation of the CS path integrals for the 
hydrogen atom and other systems with known dynamical symmetry.

\section{The hydrogen atom: $\mbox{O}(4,2)$ dynamical group and reductions to
subgroups}

Denote generators of the dynamical group $\mbox{O}(4,2)$ of the hydrogen atom 
\cite{Zverev:barut&raczka,Zverev:bechler} (the group of rotations in 
six-dimensional pseudo-Euclidean space with a metric 
$g=\mbox{diag}(1,1,1,1,-1,-1)$) as $\mathcal{L}_{\alpha \beta}$ (in the 
notation used in \cite{Zverev:zverev&rubinstein}):
\begin{eqnarray}
&&\mathcal{L}_{\alpha 0} = (\mathbf{r} \times \mathbf{p})_{\alpha}, \; 
\mathcal{L}%
_{\alpha 1} = (r_{\alpha} p^2 - 2 p_{\alpha} \mathbf{rp} + r_{\alpha})/2,
\nonumber \\
&& \mathcal{L}_{\alpha 2} = -r p_{\alpha},\; \mathcal{L}_{\alpha 3} = %
(r_{\alpha} p^2 - 2 p_{\alpha} \mathbf{rp} - r_{\alpha})/2,
\label{Zverev:eq1} \\
&&\mathcal{L}_{01} = (rp^2-r)/2,\; \mathcal{L}_{02} = \mathbf{rp} - i,\;%
 \mathcal{L}_{03} = (rp^2+r)/2.
\nonumber
\end{eqnarray}
The Schr\"odinger equation for the hydrogen atom reads $(\mathcal{H}-E)
| \Psi%
 \rangle = 0$ with $\mathcal{H} = p^2/2 - a/r$ . After multiplication by 
$r$ it can be rewritten in terms of the generators (\ref{Zverev:eq1}) 
\cite{Zverev:bechler,Zverev:fock,%
Zverev:bednar}: 
\begin{equation}
\label{Zverev:eq2}
r(\mathcal{H}-E) | \Psi \rangle = \{(\mathcal{L}_{03}+\mathcal{L}_{01})/2 - 
E(\mathcal{L}%
_{03}-\mathcal{L}_{01})-a\}| \Psi \rangle=0.
\end{equation}
The unitary transformation
$$ 
%\begin{equation*}
%\label{Zverev:eq3}
| \overline{ \Psi} \rangle=\exp(i \theta\mathcal{L}_{02})| \Psi \rangle, \ \ 
\tanh
\theta= (1+2E)/(1-2E),
%\end{equation*}
$$
reduces (\ref{Zverev:eq2}) to
$$ 
%\begin{equation*}
%\label{Zverev:eq4}
(\mathcal{L}_{03} - a/\sqrt{-2E} )| \overline{ \Psi} \rangle= 0.
%\end{equation*}
$$
The eigenstates of $\mathcal{L}_{03}$ may be chosen in this \textit{%
auxiliary representation} (AR) as $| \overline{n,l,m} \rangle_{sph}$ or $| 
\overline{n,n_1,n_2,m} \rangle_{par}$ with $n=a/\sqrt{-2E}$, in the
spherical or parabolic coordinates. An inverse transformation
$$
%\begin{equation*}
%\label{Zverev:eq5}
\left\{ 
\begin{array}{l}
| n,l,m \rangle_{sph} \\ 
| n,n_1,n_2,m \rangle_{par}
\end{array}
\right\} = \exp(-i \theta\mathcal{L}_{02}) \left\{ 
\begin{array}{l}
| \overline{n,l,m} \rangle_{sph} \\ 
| \overline{n,n_1,n_2,m} \rangle_{par}
\end{array}
\right\},
%\end{equation*}
$$
produces the \textit{physical representation} (PR) of the eigenvectors.

Consider three types of reduction to subgroups and corresponding Lie 
algebras: 
\begin{eqnarray}
(a) & \mbox{O}(4,2) \supset \mbox{O}(4) \sim \mbox{O}(3) \otimes 
\mbox{O}(3), & \mbox{o}(3) \oplus \mbox{o}(3) = \{\mathcal{P}%
_{\alpha}^{(+)}\} \oplus \{\mathcal{P}_{\alpha}^{(-)}\},
\nonumber \\
(b) & \mbox{O}(4,2)\supset \mbox{O}(2,2) \sim \mbox{O}(2,1) \otimes 
\mbox{O}(2,1), & \mbox{o}(2,1) \oplus \mbox{o}(2,1) = \{\mathcal{Q}%
_{\alpha}^{(+)}\} \oplus \{\mathcal{Q}_{\alpha}^{(-)}\},
\label{Zverev:eq6} \\
(c) & \mbox{O}(4,2) \supset \mbox{O}(3) \otimes \mbox{O}(2,1), 
& \mbox{o}(3) \oplus \mbox{o}(2,1) = \{\mathcal{L}_{\alpha 0}\} 
\oplus \{\mathcal{L}_{0 \alpha}\},
\nonumber
\end{eqnarray}
where
$$ 
%\begin{equation*}
%\label{Zverev:eq7}
\mathcal{P}_{\alpha}^{(\pm)} = (\mathcal{L}_{\alpha 0} \pm 
\mathcal{L}_{\alpha 3})/2, \ \ \ \mathcal{Q}_{\alpha}^{(\pm)} = (%
\mathcal{L}_{0 \alpha} \pm \mathcal{L}_{3 \alpha})/2.
%\end{equation*}
$$
For these cases the space $R$ of bound states of the hydrogen atom is 
decomposed as follows:
\begin{eqnarray}
(a) &&R=\oplus _{n\ge 1}\{R({}^{\mathbf{O(3)}}n-1,(n_1))\otimes R({}^{%
\mathbf{O(3)}}n-1,(n_2))\},
\label{Zverev:eq8} \\
(b) &&R=\oplus _{m\in Z}\{R({}^{\mathbf{O(2,1)}}1+|m|,(n_1+\frac
12(m-|m|)))\otimes R({}^{\mathbf{O(2,1)}}1+|m|,(n_2+\frac 12(m-|m|)))\}, 
\nonumber \\
(c) &&R=\oplus _{l\ge 0}\{R({}^{\mathbf{O(3)}}2l,(l-m))\otimes R({}^{\mathbf{%
O(2,1)}}2l+2,(n-1-l))\},
\nonumber
\end{eqnarray}
where $R({}^{\mathbf{G}}S,(K))=\left\{ |{}^{\mathbf{G}}S,(K)\rangle \right\}$
denotes the irreducible representation space of a group $G=\mbox{O}(3)$ or 
$\mbox{O}(2,1)$ with generators $\{E_{\alpha }\}$, for which we have
$$
%\begin{equation*}
%\label{Zverev:eq9}
E_3|{}^{\mathbf{G}}S,(K)\rangle = (S/2-\epsilon _{\mathbf{G}%
}K ) |{}^{\mathbf{G}}S,(K)\rangle,
%\end{equation*}
$$
$$
%\begin{equation*}
%\label{Zverev:eq10}
(E_1^2+E_2^2+\epsilon_{\mathbf{G}}E_3^2)|{}^{\mathbf{G}}S,(K)\rangle = (%
S/2-\epsilon _{\mathbf{G}}S^2/4) |{}^{\mathbf{G}}S,(K)\rangle ,
%\end{equation*}
$$
where $\epsilon_{O(3)}=1,$ $\epsilon_{O(2,1)}=-1,$
and the representation space $R({}^{\mathbf{G}}S,(K))$ is spanned by the set 
of orthonormal eigenvectors $|{}^{\mathbf{G}}S,(K)\rangle$ . The vector 
subspaces 
\begin{eqnarray*}
R_0^{[+]} &=&\{|n,n-1,n-1\rangle _{sph}\}=\{|n,0,0,n-1\rangle_{par}\},
\label{Zverev:eq11}\\
R_0^{[-]} &=&\{|n,n-1,-n+1\rangle _{sph}\}=\{|n,n-1,n-1,-n+1\rangle_{par}\},
\nonumber
\end{eqnarray*}
with maximal $m=n-1$ or minimal $m=-n+1$ values of the magnetic quantum
number correspond to the circular orbits \cite{Zverev:brown}. The vectors from 
$R_0^{[\pm ]}$ play the role of reference vectors in decompositions 
(\ref{Zverev:eq8}). On the other hand, $R_0^{[-]}$ and $R_0^{[+]}$ are
irreducible spaces of $\mbox{O}(2,1)$ ($S=1$). The corresponding $\mbox{O}(2,1)$ 
Lie algebras have the form
$$
%\begin{equation*}
%\label{Zverev:eq12}
\mathcal{H}_1^{[\pm ]}= (\pm \mathcal{L}_{11}-\mathcal{L}_{22})/2 ,\ \ \ 
\mathcal{H}_2^{[\pm ]}= (\mathcal{L}_{21}\pm \mathcal{L}_{12})/2 ,\ \ \ 
\mathcal{H}_3^{[\pm ]}= (\pm \mathcal{L}_{03}\pm \mathcal{L}_{30})/2 ,\
\ \ 
%\end{equation*}
$$
where $\pm $ corresponds to the subspace $R_0^{[\pm ]}$ with $m=\pm (n-1)$.
Following \cite{Zverev:mostowski}, we construct coherent states in two stages. 
At first, using the operators $\{\mathcal{H}_\alpha ^{[\pm ]}\}$, we construct 
the circular orbits CS in $R_0^{[\pm ]}$. Next, we construct CS in the global
space $R$ using one of the generator sets from (\ref{Zverev:eq6}). Each 
reduction scheme corresponds to a specific construction method of the hydrogen 
atom CS: 
\begin{eqnarray}
(a) &&|{}^{\mathbf{O(4)}}\omega ^{[\pm ]},\tau ,\eta \rangle =\mathcal{D}%
_{\mathbf{I}}({}^{\mathbf{O(3)}}\tau ,(\mathcal{P}_\alpha ^{(+)}))\mathcal{D}_%
{\mathbf{I}}({}^{\mathbf{O(3)}}\eta ,(\mathcal{P}_\alpha ^{(-)}))|\omega ^{%
[\pm ]}\rangle , 
\nonumber \\
(b) &&|{}^{\mathbf{O(2,2)}}\omega ^{[\pm ]},\tau ,\eta \rangle =\mathcal{D}%
_{\mathbf{I}}({}^{\mathbf{O(2,1)}}\tau ,(\mathcal{Q}_\alpha 
^{(+)}))\mathcal{D}_%
{\mathbf{I}}({}^{\mathbf{O(2,1)}}\eta ,(\mathcal{Q}_\alpha ^{(-)}))|\omega ^{%
[\pm ]}\rangle ,
 \label{Zverev:eq13} \\
(c) &&|{}^{\mathbf{O(3)\otimes O(2,1)}}\omega ^{[\pm ]},\tau ,\eta \rangle =%
\mathcal{D}_{\mathbf{I}}({}^{\mathbf{O(3)}}\tau ,(\mathcal{L}_{\alpha 0}))%
\mathcal{D}_{\mathbf{I}}({}^{\mathbf{O(2,1)}}\eta ,(\mathcal{L}_{0\alpha }))|%
\omega ^{[\pm]}\rangle ,  \nonumber
\end{eqnarray}
where 
\begin{equation}
\label{Zverev:eq14}
|\omega ^{[\pm ]}\rangle =\mathcal{D}_{\mathbf{II}}({}^{\mathbf{O(2,1)}%
}\omega ,(\mathcal{H}_\alpha ^{[\pm ]}))|1,0,0,0\rangle _{par}\;
\end{equation}
is the circular orbit CS. We use an uniform notation for the generating
operator of CS:
$$
%\begin{equation*}
%\label{Zverev:eq15}
|{}^{\mathbf{G}}S,(\omega )\rangle =\mathcal{D}_\Omega ({}^{\mathbf{G}%
}\omega ,(E_\alpha ))|{}^{\mathbf{G}}S,(0)\rangle ,
%\end{equation*}
$$
where $G=\mbox{O}(3)$ or $\mbox{O}(2,1)$, and $\Omega ={\mathbf{I}}$ or 
${\mathbf{II}}$. An important point is that one should use two types of 
generating operators of CS, with different procedures for transition to 
the classical limit:  $S\rightarrow \infty$ ($\Omega ={\mathbf{I}}$);
$|\omega |\rightarrow \infty$ ($\Omega ={\mathbf{II}}$). 

There exist many algorithms to construct coherent states of dynamical groups
and quantum physical systems (see review \cite{Zverev:zhang&feng&gilmore}).
In most cases these CS provide a transition to the "classical limit" in
some sense. However, a selection of the CS for a specific physical system must
be done with great care. It is essential to take into account not only the 
symmetry of a model, but also the type of semiclassical behavior. We consider
some special types of CS and clarify conditions of quantum-classical 
correspondence.

Usually CS $\left| \xi \right\rangle $ is called semiclassical if
an overlapping distribution function $\left| \left\langle \xi \right. \left|
\xi +\delta \xi \right\rangle \right|^2$ has a sharp peak for small $|\delta 
\xi|$, and becomes singular delta-shaped function in the classical limit, 
which can be written in the form
\begin{equation}
\label{Zverev:eq16}
\left\langle \xi \right. \left| \eta \right\rangle 
\stackrel{cl. lim.}{\longrightarrow }0, \ \xi \neq \eta; \ \left\langle \xi
\right.
\left| \xi \right\rangle =1.
\end{equation}
Satisfaction of the condition
\begin{equation}
\label{Zverev:eq17}
\left\langle \xi \right| E_\alpha ^2\left| \xi \right\rangle /\left(
\left\langle \xi \right| E_\alpha \left| \xi \right\rangle \right) ^2%
\stackrel{cl. lim.}{\longrightarrow }1,
\end{equation}
where $E_\alpha$ are elements of the Lie algebra $L$, gives another
evidence of semiclassical properties of the CS $\left| \xi \right\rangle$. 
The conditions (\ref{Zverev:eq16}), (\ref{Zverev:eq17}) are satisfied in 
the cases: the Barut-Girardello $\mbox{O}(2,1)$ CS, $\left| \xi \right|
\rightarrow \infty $, $L=\mbox{o}(2,1)$; the Perelomov $\mbox{O}(2,1)$ 
CS, $S\rightarrow \infty $, $L=\mbox{o}(2,1)$; the Perelomov $\mbox{O}(3)$ 
CS, $S\rightarrow \infty $, $L=\mbox{o}(3)$.  

An extended class of semiclassical CS follows from a definition of the 
\textit{generalized hypergeometric} CS \cite{Zverev:zverev}: 
\begin{equation}
\label{Zverev:eq18}
|\xi \rangle =N_\xi ^{-1}\sum_{n=0}^\infty \sqrt{\frac{\prod_{i=1}^p(\alpha _%
i)_n}{\prod_{j=1}^q(\rho _j)_n}}\;\frac{\xi ^n}{\sqrt{n!}}\;|n\rangle ,
\end{equation}
where $(\alpha )_n$ is the Pochhammer symbol and ${}_pF_q$ denotes the
generalized hypergeometric function; $\xi $ is a complex argument; integers
$\alpha _i$ are negative for $i=1,...,l$ and positive for
$i=l+1,...,p$; $0\leq l\leq p$; $\rho _j$ are positive real numbers.
The overlapping function for this states has the form:
$$
%\begin{equation*}
%\label{Zverev:eq19}
\langle \xi |\zeta \rangle =N_\xi ^{-1}N_\zeta ^{-1}{}_pF_q((\alpha_%
p),(\rho _q),(-1)^l\xi ^{*}\zeta ), \ \ \ \langle \xi |\xi \rangle =1.
%\end{equation*}
$$
Starting from the definition (\ref{Zverev:eq18}), let us consider a number 
of special realizations:

(i)  $p=q+1\geq 1$, $\rho _j=1$, $\alpha _j=S$, $S=1,2,...$; in the special
case $p=1$ the state (\ref{Zverev:eq18}) is the Perelomov $\mbox{O}(2,1)$ CS 
\cite{Zverev:perelomov};

(ii) $p=q+1\geq 1$, $\rho _j=1$, $\alpha _j=-S$, $S=1,2,...$; in the special
case $p=1$ the state (\ref{Zverev:eq18}) is the Perelomov $\mbox{O}(3)$ CS 
\cite{Zverev:perelomov};

(iii) $q\geq p$, $\rho _j=1$, $\alpha _j=S$, $S=1,2,...$; in the case $p=q=0$
the state (\ref{Zverev:eq18}) is the CS of a harmonic oscillator 
\cite{Zverev:klauder&glauber&sudarshan};  in the case $p=-1$, $q=0$ this is 
the Barut-Girardello CS \cite{Zverev:barut&girardello}.

It can be shown directly that for the cases (i) and (ii) the requirement
(\ref{Zverev:eq16}) is satisfied for the limiting condition 
$S\rightarrow \infty $ ($\Omega =\mathbf{I}$), while for the case
(iii) the proper condition is $\left| \xi \right|\rightarrow \infty $ 
($\Omega =\mathbf{II}$). 

For comparison, we consider briefly an alternative extension of a family of 
CS for $\mbox{O}(3)$ and $\mbox{O}(2,1)$ groups - the Brif's \textit{algebra 
eigenstates} \cite{Zverev:brif}:
$$
%\begin{equation*}
%\label{Zverev:eq20}
(\beta _1E_1+\beta _2E_2+\beta _3E_3)\left| \zeta ,\beta _1,\beta _2,\beta
_3\right\rangle =\zeta \left| \zeta ,\beta _1,\beta _2,\beta
_3\right\rangle ,
%\end{equation*}
$$
where $\left| \zeta ,\beta _1,\beta _2,\beta _3\right\rangle $ is a linear
superposition of $\left| ^{\mathbf{G}}S,(K)\right\rangle $ with different 
$K$. In the \textit{general} case $b=(\beta _1^2+\beta _2^2+\epsilon _{%
\mathbf{G}}\beta _3^2)^{1/2}\neq 0$ and for $\zeta =(l-\epsilon _{%
\mathbf{G}}S/2)b$, $l=0,1,2...$, the condition (\ref{Zverev:eq16}) is
satisfied at  $S\rightarrow \infty $ ($\Omega =\mathbf{I}$). 
At the same time, in the \textit{degenerate} case $b=0$ the $\mbox{O}(2,1)$ 
algebra eigenstates are semiclassical at the both $S\rightarrow \infty $ 
and $\left| \zeta \right| \rightarrow \infty $ limiting conditions 
($\Omega =\mathbf{I}$ and $\mathbf{II}$).

These examples do not cover all diversity of CS with semiclassical
behavior.

\section{Semiclassical asymptotics of coherent states wave functions}

We use the standard saddle-point method of asymptotic estimate for integrals
$$
%\begin{equation*}
%\label{Zverev:eq21}
F(\lambda ,\mathbf{x})=\int_c\exp \{\lambda f(\mathbf{x},\mathbf{t})\}\phi%
(\mathbf{x},\mathbf{t})d\mathbf{t},
%\end{equation*}
$$
where $\lambda \gg 1$, $\mathbf{x}=\{x_1,x_2,\ldots ,x_n\}$ and  $\mathbf{t}%
=\{t_1,t_2,...,t_m\}$;  $f(\mathbf{x},\mathbf{t})$ and $\phi (\mathbf{x},%
\mathbf{t})$ are holomorphic functions; $c$ is a $n$-dimensional smooth
manifold deformed to reach a minimax (a saddle point is
nonsingular). The asymptotic formula reads \cite{Zverev:fedorjuk}
\begin{equation}
\label{Zverev:eq22}
F(\lambda,\mathbf{x})\sim \left( 2\pi /\lambda \right) ^{m/2}\exp \{\lambda
f(\mathbf{x},\mathbf{t}_0(\mathbf{x}))\}\phi (
\mathbf{x},\mathbf{t}_0(\mathbf{x}))/\sqrt{\det
\left[ -\mu \left( \mathbf{x},\mathbf{t}_0\left( \mathbf{x}\right) \right)
\right] },
\end{equation}
where $\mathbf{t}=\mathbf{t}_0(\mathbf{x})$ is a solution of the system of
equations $\partial f/\partial t_i=0$, $i=1,...,m$, for a stationary point,
and $\mu_{ij} =\partial ^2f/\partial t_i\partial t_j$. We also
assume that $|F(\lambda,\mathbf{x})|$ has the delta-shaped peak near the point
$\mathbf{x}=\mathbf{x}_{00}$. Since 
$|F(\lambda,\mathbf{x})|\sim |\exp \{\lambda
f(\mathbf{x},\mathbf{t}_0(\mathbf{x}))\}|
\sim\exp \{\lambda \mbox{Re}f(\mathbf{x},\mathbf{t}_0(\mathbf{x}))\}$,
we can find $\mathbf{x}_{00}$ as an extremal
point of the real part of the exponent. The supplementary equations for the 
stationary point have the form 
$$
%\begin{equation*}
%\label{Zverev:eq23}
\frac{\partial f(\mathbf{x},\mathbf{t}_0(\mathbf{x}))}{\partial x_i}+c.c.=%
\frac{\partial f(\mathbf{x},\mathbf{t})}{\partial x_i}\left| _{\mathbf{t}=%
\mathbf{t}_0(\mathbf{x})}+c.c.=0.\right. 
%\end{equation*}
$$
Substituting the second order expansion of $f(\mathbf{x},\mathbf{t}_0(\mathbf{%
x}))$ at $\mathbf{x}=\mathbf{x}_{00}$ into
(\ref{Zverev:eq22}), we find the desired expression for the asymptotic estimate
\begin{eqnarray*}
F(\lambda ,\mathbf{x}) &\sim &\left( 2\pi /\lambda \right) ^{m/2}\sqrt{\det 
\left[ -\mu_{00}\right] }^{-1}\phi _{00}\exp \left\{ \lambda f_{00}+i \lambda 
\mathbf{h}^{T}\delta \mathbf{x}\right.   \nonumber \\
&&\left. +\left( \lambda/2 \right) \delta \mathbf{x}^{T}\left( \sigma
_{00}-\gamma_{00}^{T}\,\mu _{00}^{-1}\,\gamma _{00}\right) \delta
\mathbf{x}\right\},
\label{Zverev:eq24}
\end{eqnarray*}
with
\[
\sigma _{00}=\left\{ \frac{\partial ^2f_{00}}{\partial x_k\partial x_l}%
\right\} ,\qquad \gamma _{00}=\left\{ \frac{\partial ^2f_{00}}{\partial
x_l\partial t_r}\right\} ,\qquad \mathbf{h}=\left\{ \mbox{Im}\!\left[ \frac{%
\partial f_{00}}{\partial x_k}\right] \right\},
\]
where we use the notation $\left( \cdot \right) _{00}\equiv \left. \left(
\cdot \right) \right| _{x=x_{00},\,t=t_{00}}$ for different functions given
in the point $\left( \mathbf{x}_{00},\mathbf{t}_{00}\right) $, $\mathbf{%
t}_{00}\equiv \mathbf{t}_0(\mathbf{x}_{00})$.

Consider in detail a special case of hydrogenic CS wave
function. Using the formulae (\ref{Zverev:eq13}c) and (\ref{Zverev:eq14})
with generating operators for the Perelomov CS (P) and the Barut-Girardello 
CS (BG), we obtain 
$$
%\begin{equation*}
%\label{Zverev:eq25}
|{}\omega ,0,\eta \rangle =\mathcal{D}_{\mathbf{I}}({}_{P}^{\mathbf{%
O(2,1)}}\eta ,(\mathcal{L}_{0\alpha }))\mathcal{D}_{\mathbf{II}}({}_{BG%
}^{\mathbf{O(2,1)}}\omega ,(\mathcal{H}_\alpha ^{[\pm ]}))|1,0,0\rangle
_{sph}.
%\end{equation*}
$$
Noting that variation of $\tau $ in (\ref{Zverev:eq13}c) implies only
a rotation of the CS wave function without change of its shape, we can set
$\tau =0$. Using the coordinate hydrogenic wave function in the \textit{%
auxiliary representation} 
$$
%\begin{equation*}
%\label{Zverev:eq26}
\langle \mathbf{r}|\overline{n,l,m}\rangle =\frac 2{(2l+1)!}\sqrt{\frac{%
(n+l)!}{(n-l-1)!}}(2r)^le^{-r}{}_1F_1(l+1-n,2l+1,2r)Y_{lm}(\theta ,\phi ),
%\end{equation*}
$$
and in the physical representation 
$$
%\begin{equation*}
%\label{Zverev:eq27}
\langle \mathbf{r}|n,l,m\rangle =\frac 1{n^2}\langle \frac{\mathbf{r}}n|%
\overline{n,l,m}\rangle ,
%\end{equation*}
$$
we obtain the explicit form for the CS: 
$$
%\begin{equation*}
%\label{Zverev:eq28}
\langle \mathbf{r}|\overline{\omega ,0,\eta }\rangle =\frac 1{\sqrt{%
I_0(2|\omega |)}}\sum_{l=0}^\infty \sum_{n=l+1}^\infty \frac{\omega ^l}{l!}%
\frac{(1-|\eta |^2)^{l+1}}{\sqrt{(2l+1)!}}\sqrt{\frac{(n+l)!}{(n-l-1)!}}\eta
^{n-l-1}\langle \mathbf{r}|\overline{n,l,m}\rangle .
%\end{equation*}
$$
Performing the summation we arrive at 
$$
%\begin{equation*}
%\label{Zverev:eq29}
\langle \mathbf{r}|\overline{\omega ,0,\eta }\rangle =\frac 1{\sqrt{\pi
I_0(2|\omega |)}}\frac{1-|\eta |^2}{(1-\eta )^2}\exp \left( -r\frac{1+\eta }{%
1-\eta }\right) I_0\left( \frac{2\sqrt{-\omega r_{+}(1-|\eta |^2)}}{1-\eta }%
\right) ,
%\end{equation*}
$$
where $r_{+}=r\exp (i\phi ),\;r=|\mathbf{r}|$ and $I_0$ denotes the modified
Bessel function. The asymptotic estimate (obtained directly or by the
saddle-point method) for the absolute value of the CS wave function has the
form: 
$$
%\begin{equation*}
%\label{Zverev:eq30}
|\langle \mathbf{r}|\overline{\omega ,0,\eta }\rangle |^2\approx \frac
1{2\pi ^{3/2}|\omega |^{1/2}}\frac{(1-|\eta |^2)^2}{|1-\eta |^4}\exp \left\{
-\frac 1{2|\omega |}\frac{(1-|\eta |^2)^2}{|1-\eta |^4}(\Delta x^2+\Delta
y^2+2\Delta z^2),\right\} ,
%\end{equation*}
$$
where $\Delta x=x-x_0,\;\Delta y=y-y_0$; the point with coordinates 
$$
%\begin{equation*}
%\label{Zverev:eq31}
x_0=\langle n\rangle _\infty (e-\cos \theta ),\ \ y_0=\langle n\rangle
_\infty \sqrt{1-e^2}\sin \theta 
%\end{equation*}
$$
lies on the \textit{elliptic} orbit with an eccentricity $e$: 
$$
%\begin{equation*}
%\label{Zverev:eq32}
\langle n\rangle _\infty =\frac{|\omega |(1+|\eta |^2)}{1-|\eta |^2},\ \ \ e=%
\frac{2|\eta |}{1+|\eta |^2}.
%\end{equation*}
$$
In a similar way one can find the asymptotic estimate for the physical CS
wave function, but it is omitted here due to its complexity.
In this case only the saddle-point method with interchange
of summation and integration can be used. It is remarkable that in
this case the system of equations for the saddle point admits
analytical solution. In a special case of \textit{circular} orbits the
asymptotic
estimate for auxiliary and physical representations simplifies to
\begin{equation}
\label{Zverev:eq33}
\langle \mathbf{r}_0+\Delta \mathbf{r}|\overline{\omega ,0,0}\rangle \approx
\frac 1{\sqrt{2}\pi ^{3/4}|\omega |^{1/4}}\exp \left\{ i|\omega |\Delta \phi
-\frac{z^2}{2|\omega |}-\frac{|\omega |\Delta \phi ^2}4-\frac{\Delta \rho ^2%
}{4|\omega |}+\frac{i\Delta \rho \Delta \phi }2\right\} ,
\end{equation}
where $\mathbf{r}=(\rho \cos \phi ,\rho \sin \phi ,z),\Delta \phi =\phi
-\phi _0,\Delta \rho =\rho -|\omega |$, and 
\begin{equation}
\label{Zverev:eq34}
\langle \mathbf{r}_0+\Delta \mathbf{r}|\omega ,0,0\rangle \approx \frac 1{%
\sqrt{5}\pi ^{3/4}|\omega |^{9/4}}\exp \left\{ i|\omega |\Delta \phi -\frac{%
z^2}{2|\omega |^3}-\frac{|\omega |\Delta \phi ^2}{10}-\frac{\Delta \rho ^2}{%
10|\omega |^3}+\frac{2i\Delta \rho \Delta \phi }{5|\omega |}\right\} ,
\end{equation}
where $\mathbf{r}=(\rho \cos \phi ,\rho \sin \phi ,z),\Delta \phi =\phi
-\phi _0,\Delta \rho =\rho -|\omega |^2$. The functions (\ref{Zverev:eq33}) 
and (\ref{Zverev:eq34}) are related by the following equation: 
$$
%\begin{equation*}
%\label{Zverev:eq35}
\langle \mathbf{r}_1|\omega ,0,0\rangle =\int \frac{d\mathbf{r}_2}{r_2}K(%
\mathbf{r}_1,\mathbf{r}_2)\langle \mathbf{r}_2|\overline{\omega ,0,0}\rangle,
%\end{equation*}
$$
where 
$$
%\begin{equation*}
%\label{Zverev:eq36}
K(\mathbf{r}_1,\mathbf{r}_2)\approx \frac 1{\sqrt{10\pi ^3\rho _2^5}}\exp
\left\{ i\rho _2\Delta \phi -\frac{z_2^2}{2\rho _2}-\frac{z_1^2}{2\rho _2^3}-%
\frac{\rho _2\Delta \phi ^2}{10}-\frac{\Delta \rho ^2}{10\rho _2^3}+\frac{%
2i\Delta \rho \Delta \phi }{5\rho _2}\right\} ,
%\end{equation*}
$$
and $\Delta \phi =\phi _1-\phi _2,\Delta \rho =\rho _1-\rho _2^2$.

% \LastPageEnding{Zverev}

\end{document}